\documentclass[aps,onecolumn,preprintnumbers,nofootinbib,superscriptaddress,11pt]{revtex4}
\usepackage{amsmath} \usepackage{graphicx} \usepackage{amsfonts}
\usepackage{array} \usepackage{amsthm} \usepackage{bm}
\usepackage{palatino} \usepackage{mathpazo} 
\usepackage{supertabular}
\usepackage[breaklinks]{hyperref}
\usepackage{color}
\usepackage{epsfig}
\usepackage{epstopdf}

\newcommand{\nn}{\nonumber}
\newcommand{\be}{\begin{equation}}
\newcommand{\ee}{\end{equation}}
\newcommand{\ba}{\begin{eqnarray}}
\newcommand{\ea}{\end{eqnarray}}
\newcommand{\bal}{\begin{align}}
\newcommand{\eal}{\end{align}}

\newcommand{\ga}{\gamma}

\newcommand{\bw}{\begin{widetext}}
\newcommand{\ew}{\end{widetext}}

\begin{document}

\title{Spherical accretion by normal and phantom Einstein--Maxwell--dilaton black holes}

\author{Mustapha Azreg-A\"{\i}nou}\email{azreg@baskent.edu.tr}
\affiliation{Engineering Faculty, Ba\c{s}kent University, Ba\u{g}l\i ca Campus, Ankara, Turkey}
\author{Ayyesha K. Ahmed}\email{ayyesha.kanwal@sns.nust.edu.pk}
\affiliation{Department of Mathematics, School of Natural
	Sciences (SNS), National University of Sciences and Technology
	(NUST), H-12, Islamabad, Pakistan}
\author{Mubasher Jamil}\email{mjamil@sns.nust.edu.pk}
\affiliation{Department of Mathematics, School of Natural
	Sciences (SNS), National University of Sciences and Technology
	(NUST), H-12, Islamabad, Pakistan}

\begin{abstract}
We investigate the spherical accretion process for general static spherically symmetric fluids. We analyze this process by using the general metric ansatz for spherically symmetric black holes. We specialize to the case of normal and phantom isothermal fluids and investigate their accretion process onto normal and phantom Einstein-Maxwell-dilaton black holes. Backreaction effects are discussed.
\end{abstract}


\maketitle

\section{Introduction}
One of the most important discoveries in cosmology is the fact that the Universe is expanding at an accelerating rate. Evidence of this accelerating expansion has been observed by the astronomical observations such as Cosmic Microwave Background (CMB) radiation \cite{1}, Supernova Type Ia \cite{2} and large scale structure data \cite{3}. According to Einstein's theory of general relativity and Friedmann cosmology, there exists some matter with a negative pressure and its absolute value is somehow comparable to the energy density (with standard units $c=G=1$) \cite{Babichev}. In fact this matter is called dark energy and is being considered as the mainstream to the modifications of Einstein's theory as it will lead to the violation of the Weak Equivalence Principle. However, the physical origin of this energy is still unknown but many astronomical observations suggest that two thirds of the whole energy of the Universe belongs to the dark energy. We model this dark energy using a perfect fluid with the equation of state $p=k e$ where $k$ is the state parameter ($k=-1$) and $e$ is the energy density.

In the literature there is a huge list of proposed models for dark energy. According to recent observations the cosmological term requires very minute value of the energy density in the vacuum which eventually demands a very large parameter to fit in the field theory. For this reason many other models with $k\neq-1$ have been proposed. For instance, we have phantom energy: a matter whose state parameter is negative ($k<-1$) \cite{4,5,6}. Different aspects of phantom cosmology have already been discussed in the literature \cite{7,8}. One of the exceptional hypothesis in cosmology is the Big Rip scenario \cite{5} which predicts that due to the expansion of the Universe all the matter will tear apart at a finite time interval. Hence this scenario could also justify another form of energy called phantom energy.

Accretion is a process by which a gravitating object such as a black hole or a massive star can capture particles from its vicinity which eventually leads to a change in the physical properties of the accreting body \cite{9,10,11,12}. It is considered as one of the most pervasive process in the Universe. In fact the supermassive black holes at the center of galaxies suggest that black holes could have evolved through the accretion process. However, an accretion process does not always increase the mass of the compact object but it could decrease it (a) when the infalling matter is thrown out in the form of jets or cosmic rays or if the infalling gas is a phantom matter~\cite{Babichev2,Jamil}, or (b) if the compact object is a superspinar, say a naked singularity, where the decrease of mass may possibly occur when matter from orbits with negative energy plunges into the singularity as was noticed for the Kerr naked singularity~\cite{super1,super2}. In the literature there is not any substantial  contribution regarding exotic matter. Since our Universe is highly dominated by dark energy and phantom energy therefore, it is more interesting to study the accretion of such energies onto black holes and the accretion of ordinary and phantom matter onto phatom black holes.

The first study on accretion process in the Newtonian framework was done by Bondi in 1952 \cite{Bondi}. Through the evolution of Einstein's theory of gravitation, Michel was the one who investigated the gas accretion onto Schwarzschild black hole in relativistic framework \cite{Michel}. Babichev et al. discussed the stationary black hole in the phantom energy dominated Universe \cite{Babichev}. They have found that phantom energy and dark energy will decrease the mass of the black hole. The effects of phantom accretion and Chaplygin gas were investigated onto the charged black hole by Jamil et al. \cite{Jamil}. Debnath further generalizes the above idea and presented a general framework of a static accretion process onto static, spherically symmetric black holes \cite{Debnath}. Moreover, the accretion process of a spherically symmetric spacetime is investigated in a series of our recent papers \cite{A1,A2,A3,A4}. Accretion of rotating fluids onto stationary solutions has been investigated by one of us~\cite{A5}.

In Ref. \cite{13}, the dynamical behavior of phantom energy near a five-dimensional charged black hole has been considered. The authors formulated the equations for steady state, spherically symmetric flow of phantom fluids onto the black hole and concluded that a five-dimensional black hole cannot be transformed into an extremal black hole. In Ref. \cite {14}, it was shown that the size of the black hole decreases due to the phantom energy accretion. Amani et al. investigated the phantom energy accretion onto the Schwarzschild anti de-Sitter black hole with topological defect \cite{Amani}. In this work we will develop a general formalism of spherical accretion of ordinay and phantom matter by ordinay and phantom black holes.

We use the standard geometric units ($c=G=1$) and the chosen metric signature is $(+,-,-,-)$.  This paper is organized as follows: In Sec.~\ref{secge} we develop a general formalism for spherical accretion that applies to all static black holes. In Sec.~\ref{sechs} we derive the Hamiltonian system and in Sec.~\ref{seccp} we determine the critical points (CPs). In Sec.~\ref{secemd} we apply our results to the Einstein--Maxwell--dilaton (EMD) black holes. In Sec.~\ref{secif} we specialize to a particular type of fluids i.e. \emph{ordinary} and \emph{phantom} isothermal test fluids and apply our formalism to the generic case as well as to the special cases of ultra-stiff, ultra-relativistic, radiation, and sub-relativistic fluids.  In Sec.~\ref{secmar} we discuss the associated mass accretion rate and backreaction. We conclude in Sec.~\ref{seccon}. An Appendix section has been added to complete the discussion of Sec.~\ref{secif}.

\section{General equations\label{secge}}
In this section, we consider a general metric ansatz of the form
\begin{eqnarray}\label{1}
ds^{2}&=&A(r)dt^{2}-\frac{dr^{2}}{B(r)}-C(r)(d\theta^{2}+\sin^{2}\theta d\phi^{2}),
\end{eqnarray}
whose determinant is given by $g=-AC^{2}/B$. Our results will remain valid for all solutions having the line element of the form \eqref{1} and this includes all known static black holes, wormholes and other relevant solutions. 

Now we define the governing equations which we need for the process of spherical accretion. For this, we have two basic conservation laws i.e. particle number conservation and energy conservation. We consider the flow of a perfect fluid onto black hole. The four velocity of the particles is given by $u^{\mu}=\frac{dx^{\mu}}{d\tau}$. If $n$ is the particle's density, then the particle's flux is given by $J^{\mu}=nu^{\mu}$. The law of particle conservation states that there will be no change in the number of particles; particles can never be created nor destroyed i.e. the current density of the particles is divergence free,
\begin{eqnarray}\label{2}
\nabla_{\mu}J^{\mu}&=&\nabla_{\mu}(nu^{\mu})=0,
\end{eqnarray}
where $\nabla_{\mu}$ shows the covariant derivative. The energy momentum tensor for a perfect fluid is given by
\begin{eqnarray}\label{3}
T^{\mu\nu}&=&(e+p)u^{\mu}u^{\nu}-pg^{\mu\nu},
\end{eqnarray}
where $e$ and $p$ denotes the energy density and pressure respectively. The law of energy conservation shows that there will be no change in the total energy of the system and is given by
\begin{eqnarray}\label{4}
\nabla_{\mu}T^{\mu\nu}&=&0.
\end{eqnarray}
On the equatorial plane $(\theta=\pi/2)$, the continuity equation given by \ref{2} yields
\begin{equation}\label{5}
C\sqrt{D}nu=C_{1}\quad\text{with}\quad D\equiv \frac{A}{B},
\end{equation}
where $C_{1}$ is a constant of integration, which is negative for an accretion process and positive otherwise, and $u\equiv u^r$. Since the fluid is flowing radially in the equatorial plane therefore both $u^{\theta}$ and $u^{\phi}$ vanishes, and we are only left with $u^{t}$ and $u^{r}=u$ components. By using the normalization condition $(g_{\mu\nu}u^{\nu}u^{\nu}=1)$ we have
\begin{eqnarray}\label{6}
u^{t}&=&\pm \sqrt{\frac{1+B^{-1}u^{2}}{A}},
\end{eqnarray}
where the minus sign corresponds to accretion. Consequently, we obtain
\begin{eqnarray}\label{7}
u_{t}&=&g_{tt}u^{t}=\pm \sqrt{A(1+B^{-1}u^{2})}= \pm \sqrt{A+Du^2}.
\end{eqnarray}
As we have considered the steady state and spherically symmetric line element, so all the physical parameters e.g. particles density, energy density, pressure, four velocity etc. are functions of the radial coordinate $r$ only \cite{A1}.

\par The thermodynamics of the fluid is described by
\begin{eqnarray}\label{8}
dp=n(dh-Tds),\qquad de=hdn+nTds,
\end{eqnarray}
where $T$ is the temperature, $s$ is the entropy and $h$ is the specific enthalpy (enthalpy per particle) given by
\begin{equation}\label{9}
h=\frac{e+p}{n}.
\end{equation}
Since ordinary matter satisfies the constarint $e+p>0$, we have $h>0$ while phantom matter violates it; that is, $e+p<0$ for phantom matter resulting in $h<0$.

In relativistic hydrodynamics, there exists a scalar $hu_{\mu}\xi^{\mu}$ which is conserved along the trajectories of the fluid \cite{Rezolla} i.e.
\begin{eqnarray}\label{10}
u^{\nu}\nabla_{\nu}(hu_{\mu}\xi^{\mu})&=&0,
\end{eqnarray}
where $\xi^{\mu}$ is Killing vector of the spacetime. For instance, if we take $\xi^{\mu}=(1,0,0,0)$ we obtain~\cite{A3}
\begin{eqnarray}\label{11}
\partial_{r}(hu_{t})&=&0~~~~~\text{or}~~~~~h\sqrt{A+Du^2}=C_{2},
\end{eqnarray}
where $C_{2}$ is a constant of integration: $C_{2}>0$ for ordinary matter and $C_{2}<0$ for phantom matter. 

Since in the equatorial plane $(\theta=\pi/2)$ the motion is radial, $d\theta=d\phi=0$ and so we can decompose our metric (\ref{1}) as
\begin{eqnarray}\label{12}
ds^{2}=\Big(\sqrt{A}dt^{2}\Big)^{2}-\Big(\sqrt{\frac{1}{B}}dr^{2}\Big)^{2},
\end{eqnarray}
in the standard relativistic way \cite{b1,b2} as seen by a local static observer. We can define the three velocity by
\begin{eqnarray}\label{13}
v&=&\sqrt{\frac{1}{AB}}\frac{dr}{dt}.
\end{eqnarray}
This leads to
\begin{eqnarray}\label{14}
v^{2}&=&\frac{1}{AB}\Big(\frac{u}{u^t}\Big)^{2},
\end{eqnarray}
where $u^{t}=dt/d\tau$ and $u^{r}=u=dr/d\tau$. Using (\ref{6}) and isolating $u^{2}$ we obtain~\cite{A3}
\begin{eqnarray}
\label{15a}u^{2}&=&\frac{Bv^{2}}{1-v^{2}},\\
\label{15}u_{t}^{2}&=&\frac{A}{1-v^{2}},\\
\label{15b}v^2&=&\frac{Du^2}{A+Du^2},
\end{eqnarray}
and hence Eq. (\ref{5}) becomes~\cite{A3}
\begin{eqnarray}\label{16}
\frac{A(Cnv)^{2}}{1-v^{2}}&=&C_{1}^{2}.
\end{eqnarray}
We use these results below in the Hamiltonian analysis.

Based solely on~\eqref{16}, it was concluded in Ref.~\cite{A3} that the behavior of the fluid near the horizon is independent on the form of $A(r)$. More precisely, the fluid reaches the horizon either with $v\to 0$ or $v\to 1$~\cite{A1,A2}.

The constant $C_1^2$ in~\eqref{16} can be written as $A_0C_0^2n_0^2v_0^2/(1-v_0^2)$ where {\textquotedblleft 0\textquotedblright} denotes any reference point ($r_0,\,v_0$) from the phase portrait; this could be a CP, if there is any, spatial infinity ($r_{\infty},\,v_{\infty}$), or any other reference point. We can thus write~\cite{A3}
\begin{equation}\label{6b}
\frac{n^2}{n_0^2}=\frac{A_0C_0^2v_0^2}{1-v_0^2}~\frac{1-v^2}{AC^2v^2}=\frac{C_1^2}{n_0^2}~\frac{1-v^2}{AC^2v^2}.
\end{equation}

\section{Hamiltonian system\label{sechs}}
We have two integrals of motion $C_{1}$ and $C_{2}$ given by Eqs. (\ref{5}) and (\ref{11}) respectively. By Bernoulli's theorem, the square of $C_{2}$ in Eq. (\ref{11}) is proportional to the fluid energy. So by using (\ref{15}) we can define the Hamiltonian as~\cite{A3}
\begin{eqnarray}\label{17}
\mathcal{H}(r,v)&=&\frac{h^{2}(r,v)A(r)}{1-v^{2}},
\end{eqnarray}
where $v$ is the three velocity given by Eq. (\ref{15b}).

\section{Critical points\label{seccp}}
It is well known that a perfect fluid~\eqref{3} is adiabatic; that is, the specific entropy is conserved along the evolution lines of the fluid: $u^{\mu}\nabla_{\mu}s=0$ (See~\cite{A1} for a proof). In the special case we are considering in this work where the fluid motion is radial, stationary (no dependence on time), and it conserves the spherical symmetry of the black hole, the latter equation reduces to $\partial_rs=0$ everywhere, that is, $s\equiv \text{constant}$. Thus, the motion of the fluid is isentropic and equations~\eqref{8} reduce to
\begin{eqnarray}\label{18}
dp&=&ndh,~~~~~~~~~~de=hdn.
\end{eqnarray}
The adiabatic speed of sound is defined by
\begin{eqnarray}\label{19}
a^{2}&=&\frac{dp}{de}=\frac{d\ln h}{d\ln n}.
\end{eqnarray}
Now, with $\mathcal{H}$ given by Eq. (\ref{17}), the dynamical system reads
\begin{eqnarray}\label{20}
\dot{r}&=&\mathcal{H}_{,v},~~~~~~~~~~\dot{v}=-\mathcal{H}_{,r},
\end{eqnarray}
where the dot denotes the $\bar{t}$ derivative where $\bar{t}$ is the time variable of the Hamiltonian dynamical system. Using the results of Ref.~\cite{A3} we obtain
\begin{align}
\label{d3}&\dot{r}=\frac{2h^2A}{v(1-v^2)^2}~(v^2-a^2),\\
\label{d4}&\dot{v}=-\frac{h^2}{1-v^2}\Big[\frac{d A}{d r}-2a^2A~\frac{d \ln(\sqrt{A}C)}{d r}\Big].
\end{align}
Introducing the notation $g_c=g(r)|_{r=r_c}$ and $g_{c,r_c}=g_{,r}|_{r=r_c}$ where $g$ is any function of $r$, the following equations provide a set of CPs that are solutions to $\dot{r}=0$ and $\dot{v}=0$:
\begin{equation}\label{24}
v_c^2=a_c^2\qquad\text{ and }\qquad a_c^2=\frac{C_cA_{c,r_c}}{C_cA_{c,r_c}+2AC_{c,r_c}}=\frac{C^2A_{,r}}{(C^2A)_{,r}}\Big|_{r=r_c},
\end{equation}
where $a_c$ is the three-dimensional speed of sound evaluated at the CP. The first equation states that at a CP the three-velocity of the fluid equals the speed of sound. The second equation determines $r_c$ once an equation of state is known.

\section{Spherical accretion by Einstein--Maxwell--dilaton black holes\label{secemd}}
The Lagrangian for EMD theory \cite{d2,c1,c2} is given by
\begin{eqnarray}\label{27}
\mathcal{L}&=&R-2\eta_{1}g^{\mu\nu}\phi_{,\mu}\phi_{,\nu}+\eta_{2}e^{2\lambda\phi}F_{\mu\nu}F^{\mu\nu},
\end{eqnarray}
where $R$ is the Ricci scalar, $F_{\mu\nu}$ is the Maxwell tensor of the electromagnetic field and $\phi$ represents the dilaton field. The parameter $\lambda$ is the
real dilaton-Maxwell coupling constant, and $\eta_1=\pm 1$, $\eta_2=\pm 1$. Normal EMD corresponds to $\eta_2=\eta_1=+1$, while phantom couplings of the dilaton field $\phi$ or/and Maxwell field $F = d A$ are obtained for $\eta_1=-1$ or/and $\eta_2=-1$ yielding the theories EM$\bar{\text{D}}$ ($\eta_2=+1,\,\eta_1=-1$), E$\bar{\text{M}}$D ($\eta_2=-1,\,\eta_1=+1$), and E$\bar{\text{M}}\bar{\text{D}}$ ($\eta_2=-1,\,\eta_1=-1$). For short we call all these theories EMD theory. 

A class of spherically symmetric solutions to the field equations associated with the Lagrangian~\eqref{27} are known in the literature~\cite{d2} and are given by
\begin{eqnarray}\label{28}
ds^{2}&=&f_{+}f_{-}^{\gamma}dt^{2}-\frac{dr^{2}}{f_{+}f_{-}^{\gamma}}-r^{2}f_{-}^{1-\gamma}(d\theta^{2}+\sin^{2}\theta d\phi^{2}),
\end{eqnarray}
where we identify ($A,\,B,\,C$)~\eqref{1} by
\begin{equation}\label{id}
A=B=f_{+}f_{-}^{\gamma},\quad C=r^{2}f_{-}^{1-\gamma}.
\end{equation}
The electric and dilaton fields are given by
\begin{eqnarray}\label{29}
F&=&-\frac{Q}{r^2}dr\wedge dt,~~~~~~~~~~e^{-2\lambda\phi}=f_{-}^{1-\gamma},
\end{eqnarray}
whereas,
\begin{eqnarray}\label{30}
f_{\pm}&=&1-\frac{r_{\pm}}{r},~~~~~~~~~~\gamma=\frac{1-\eta_{1}\lambda^{2}}{1+\eta_{1}\lambda^{2}},~~~~~~~~~~\eta_1\lambda^2=\frac{1-\gamma}{1+\gamma},
\end{eqnarray}
and
\begin{equation}\label{30b}
r_+=M+M\sqrt{1-\frac{2\eta_2Q^2}{M^2}~\frac{\ga}{1+\ga}},\qquad r_-=\frac{2\eta_2Q^2}{(1+\ga)r_+}=\frac{M}{\ga}-\frac{M}{\ga}\sqrt{1-\frac{2\eta_2Q^2}{M^2}~\frac{\ga}{1+\ga}},
\end{equation}
where we have used the notation of Refs.~\cite{c1,d3}. Note that~\cite{d3}
\begin{align}
&\gamma \in (-\infty,-1)\cup[1,+\infty) \quad \text{if}\quad \eta_{1}=-1,\nn\\
\label{gamma}&\gamma \in (-1,+1] \quad\text{if}\quad \eta_{1}=+1.
\end{align}
A useful expression for our further investigations is the equation of which $r_+$ is a solution:
\begin{equation}\label{rp}
r_+^2=2Mr_+-\frac{2\eta_2Q^2\gamma}{1+\gamma}.
\end{equation}

The case $\ga=1$ corresponds to normal and phantom Reissner-Nordstr\"om black holes.

A global flow is possible if the fluid elements can reach spatial infinity. Using the expressions~\eqref{id} of $A$ and $C$ in~\eqref{16}, we see that as $r\to\infty$, $nv$ behaves as
\begin{equation}\label{gf1}
nv\sim\frac{1}{r^{2}}\qquad (\text{for all }\gamma).
\end{equation}
As $r\to\infty$, we may distinguish two cases:
\begin{align}
\label{gf2}&\text{(a)}\quad v=v_{\infty}\Big(1-\frac{v_{x_a}}{r^{x_a}}+\cdots\Big)\qquad (\text{with independent term}),\\
\label{gf3}&\text{(b)}\quad v=\frac{v_y}{r^y}\Big(1-\frac{v_{x_b}}{r^{x_b}}+\cdots\Big)\qquad (\text{no independent term}),
\end{align}
where ($v_{\infty},\,v_{x_a},\,v_{x_b},\,v_y$) are constants and $0<y\leq 2$. The corresponding expansions for $n$ are of the form
\begin{align}
\label{gf4}&\text{(a)}\quad n=\frac{n_2}{r^2}\Big(1+\frac{n_{z_a}}{r^{z_a}}+\cdots\Big),\\
\label{gf5}&\text{(b)}\quad n=\frac{n_{2-y}}{r^{2-y}}\Big(1+\frac{n_{z_b}}{r^{z_b}}+\cdots\Big),
\end{align}
where ($n_2,\,n_{z_a},\,n_{z_b},\,n_{2-y}$) are constants. Now, since the series expansion in powers of $1/r$ of the expression $AC^2/r^4$~\eqref{id} has only positive integer powers, we conlude from~\eqref{16} that the exponents (${x_a},\,{x_b},\,{z_a},\,{z_b},\,2y$), and all the exponents inside the parenthesis in~\eqref{gf2} to~\eqref{gf5}, must be positive integers too:
\begin{equation}
({x_a},\,{x_b},\,{z_a},\,{z_b},\,2y)\in \mathbb{N^+}^5.
\end{equation}
Since $0<y\leq 2$, this constraints $y$ to assume the four values
\begin{equation}
y=\frac{1}{2},\,1,\,\frac{3}{2},\,2.
\end{equation}
The case $y=1/2$ corresponds to Keplerian motion.

Substituting~\eqref{gf2} and~\eqref{gf3} into~\eqref{6b} we obtain
\begin{align}
\label{gf6}&\text{(a)}\quad n_2=\frac{C_1\sqrt{1-v_{\infty}^2}}{v_{\infty}},\quad z_a=1,\quad n_{z_a}=\left\{\begin{array}{ll}
M+(1-\gamma)r_-+\frac{v_{x_a}}{1-v_{\infty}^2}, & \hbox{$x_a=1$;} \\
M+(1-\gamma)r_-, & \hbox{$x_a\geq 2$,}
\end{array}
\right. \\
\label{gf7}&\text{(b)}\quad n_{2-y}=\frac{C_1}{v_y},\quad z_b=1,\quad n_{z_b}=\left\{\begin{array}{llll}
M+(1-\gamma)r_-+v_{x_b}-\frac{v_y^2}{2}, & \hbox{$x_b=1,\,y=\frac{1}{2}$;} \\
M+(1-\gamma)r_--\frac{v_y^2}{2}, & \hbox{$x_b\geq 2,\,y=\frac{1}{2}$;} \\
M+(1-\gamma)r_-+v_{x_b}, & \hbox{$x_b=1,\,y\geq 1$;} \\
M+(1-\gamma)r_-, & \hbox{$x_b\geq 2,\,y\geq 1$,}
\end{array}
\right.
\end{align}
where $C_1/v_{\infty}>0$ and $C_1/v_y>0$. Similarly, from the facts that $\mathcal{H}$~\eqref{17} is a constant of motion and that the series expansion of $A/(1-v^2)$ as $r\to\infty$ includes only positive integer powers of $1/r$, we conclude that the series expasion of $h$ too has only positive integer powers. Hence, we write
\begin{equation}\label{gf8}
h=h_{\infty}\Big(1+\frac{h_1}{r}+\cdots\Big),
\end{equation}
resulting, upon substituting into~\eqref{17}, in
\begin{align}
\label{gf9}&\text{(a)}\quad h_1=\left\{\begin{array}{ll}
M+\frac{v_{x_a}v_{\infty}^2}{1-v_{\infty}^2}, & \hbox{$x_a=1$;} \\
M, & \hbox{$x_a\geq 2$,}
\end{array}
\right. \\
\label{gf10}&\text{(b)}\quad h_1=\left\{\begin{array}{ll}
M-\frac{v_y^2}{2}, & \hbox{$y=\frac{1}{2}$;} \\
M, & \hbox{$y\geq 1$.}
\end{array}
\right.
\end{align}
For ordinary matter $h_{\infty}$ is the baryonic mass $m$. A further discussion of the relationships between the different parameters, in the above expressions for ($v,\,n,\,h$), would depend on an equation of state that relates $e$ to $p$~\eqref{3}. As to $n_2$ and $n_{2-y}$, they depend on the nature of the fluid where in many astrophysical applications it is taken as a perfect gas~\cite{inviscid}.

\section{Isothermal-like test fluids\label{secif}}
Isothermal flow of ordinary fluids refers to flow at constant temperature. This model of flow is a generalizarion of the classical formula $p=\rho RT$ where the fluid is assumed to be a perfect gas. Their equation of state is such that the pressure is directly proportional to the energy density: $p=k e$. This model, however, is not suitable for an isothermal flow  description of phantom fluids where $e+p<0$, resulting in $k<-1$ and $a^{2}=dp/de=k<0$.

Instead of a direct law of proportionality, we rather assume a linear dependence of $p$ and $e$~\cite{Babichev2}
\begin{equation}\label{if1}
p=\omega (e-e_0),
\end{equation}
where $0<\omega\leq 1$ and $e_0$ are constants. For ordinary fluids we may take $e_0\equiv 0$. For phantom fluids $0<e<e_0$ to ensure that $p<0$; this, however, does not ensure that $e+p<0$, so we assume that
\begin{equation}\label{if2}
e<\frac{\omega}{1+\omega}~e_0,
\end{equation}
where the rhs constitutes an upper limit for the energy density of phantom fluids.

Using~\eqref{19}, we obtain
\begin{eqnarray}\label{26a}
a^{2}=\frac{d\ln h}{d\ln n}=\omega,
\end{eqnarray}
hence
\begin{eqnarray}\label{26b}
h=\text{constant }\times n^{\omega}=\frac{h_{\infty}}{n_{\infty}^{\omega}}~n^{\omega}.
\end{eqnarray}
Now, using~\eqref{6b} and~\eqref{26b} the Hamiltonian~\eqref{17} reduces to~\cite{A3}
\begin{eqnarray}\label{26c}
\mathcal{H}(r,v)&=&\frac{1}{[vC(r)]^{2\omega}}\Big[\frac{A(r)}{1-v^{2}}\Big]^{1-\omega},
\end{eqnarray}
where all factor constants have been removed.

Restricting ourselves to isothermal fluids, the Hamiltonian~(\ref{26c}) for the phantom black hole reduces to
\begin{equation}\label{35}
\mathcal{H}=\frac{(r-r_+)^{1-\omega}(r-r_-)^{\gamma(1+\omega)-2\omega}}{v^{2\omega}r^{(1+\gamma)(1+\omega)}}.
\end{equation}
Implicit solutions to~\eqref{35}, that is, solutions to $\mathcal{H}=\text{constant}$ providing the profile of $v$ versus $r$ are similar to the plots depicted in Figs. 1 and 2 of Ref.~\cite{A1} and Figs. 1 to 4 of Ref.~\cite{A2}, and they will not be produced here. Our main purpose is to demonstrate and depict the new features pertaining to accretion onto EMD black holes.

As we mentioned above, a global flow is possible if the fluid elements can reach spatial infinity. By the last equation we see that as $r\to\infty$
\begin{equation}\label{35b}
\mathcal{H}\sim\frac{1}{v^{2\omega}r^{4\omega}},
\end{equation}
in a way independent of the value of $\gamma$. Since $\mathcal{H}\propto C_2^2$~\eqref{11} is a constant of motion, the three-velocity must behave as
\begin{equation}\label{35c}
v\sim\frac{1}{r^{2}}\qquad (\text{for all }\gamma),
\end{equation}
in the limit $r\to\infty$. As we noticed earlier, the fluid approaches the horizon in a way independent of the form of the metric component $A(r)$, and particularly, of the value of $\gamma$. Thus, the end-behavior (near the horizon or at spatial infinity) of the fluid flow does not dependent on $\gamma$ but the detailed motion of the fluid and the CPs do depend on $\gamma$.

The expression~\eqref{35c} is of the form~\eqref{gf3} with $y=2$ which means that isothermal-like fluids do not follow a Keplerian motion. Replacing $v_y$ by $v_2$ and $n_{2-y}$ by $n_{\infty}$, the expressions of ($v,\,n,\,h$) reduce to
\begin{align}
\label{gi1}&v=\frac{v_2}{r^2}\Big(1-\frac{v_{x_b}}{r^{x_b}}+\cdots\Big),\\
\label{gi2}&n=n_{\infty}\Big(1+\frac{n_{z_b}}{r}+\cdots\Big)=\frac{C_1}{v_2}\Big(1+\frac{M+(1-\gamma)r_-+\delta^1_{x_b}v_{x_b}}{r}+\cdots\Big),\\
\label{gi3}&h=h_{\infty}\Big(1+\frac{M}{r}+\cdots\Big),
\end{align}
where $C_1/v_2>0$ and
\begin{equation}\label{gi4}
M=\omega [M+(1-\gamma)r_-+\delta^1_{x_b}v_{x_b}],
\end{equation}
which results upon substituting~\eqref{gi2} and~\eqref{gi3} into~\eqref{26b}. Here $\delta^1_{x_b}$ is 1 if $x_b=1$ and 0 otherwise.

The expression for the sound speed at the CP~(\ref{24}) leads to the following expression
\begin{equation}\label{34a}
\omega =\frac{(r_+ +\gamma  r_-) r_c-(1+\gamma ) r_- r_+}{4 r_c^2-[3 r_+ +(2+\gamma ) r_-] r_c+(1+\gamma ) r_- r_+}.
\end{equation}
Using~\eqref{30b} and the third expression in~\eqref{30}, we obtain
\begin{align*}
&r_+ +\gamma  r_-=2 M,\qquad (1+\gamma ) r_- r_+=2\eta_2 Q^2,\\
&3 r_+ +(2+\gamma ) r_-=3 (r_++\gamma  r_-)+2 (1-\gamma ) r_-=6 M+\frac{4 \eta_1\eta_2\lambda ^2 Q^2}{r_+}>0.
\end{align*}
Substituting these three relations into~\eqref{34a} we bring it to the form
\begin{equation}\label{34b}
2 \omega  r_+ r_c^2-[(1+3 \omega ) M r_+ +2\eta_1\eta_2 Q^2 \lambda ^2 \omega ] r_c+\eta_2(1+\omega ) Q^2 r_+=0,
\end{equation}
yielding
\begin{equation}\label{34c}
r_c=\frac{(1+3 \omega ) M r_+ +2\eta_1\eta_2 Q^2 \lambda ^2 \omega +\sqrt{[(1+3 \omega ) M r_+ +2\eta_1\eta_2 Q^2 \lambda ^2 \omega]^2- 8\eta_2 \omega  (1+\omega ) Q^2 r_+^2}}{4 \omega  r_+}.
\end{equation}
Here $r_+$ is given by~\eqref{30b} and $\eta_1\lambda^2$ by~\eqref{30}. The other root to~\eqref{34b}, $\bar{r}_c$, is given by a similar expression to~\eqref{34c} with the minus sign in front of the square root. When the two roots are real, $r_c\geq\bar{r}_c$. In the Appendix, we will show that the roots ($r_c,\,\bar{r}_c$) are always real in the \emph{physical case} $M^2\geq Q^2$ for all $0<\omega\leq 1$ and for all values of ($M,\,Q,\,\gamma,\,\eta_2$) that make $r_+$ real.

However, a CP, yielding a critical behavior, exists only if $r_c>r_+$. With five parameters ($M,\,Q,\,\gamma,\,\eta_2,\,\omega$) being free, it is very cumbersome, even in the physical case $M^2\geq Q^2$, to compare the expression~\eqref{34c} of $r_c$ to that of $r_+$~\eqref{30b}. For that purpose we rather follow another path.

\subsection{Physical case \pmb{$M^2\geq Q^2$}}
Let us first determine a power series for $r_c$ around $\omega=1$. Note that in the case $\omega= 1$, $r_+$ is a solution to~\eqref{34b}. In fact, setting $\omega=1$ and replacing $r_c$ by $r_+$ in~\eqref{34b}, we obtain
\begin{equation}\label{pc1}
r_+^2-2 M r_+ -\eta_1\eta_2 Q^2 \lambda ^2+\eta_2 Q^2=0,
\end{equation}
where the l.h.s is zero by~\eqref{rp} and~\eqref{30}. We can determine the power series upon setting $r_c=r_++\alpha$ and $\omega=1-\epsilon$ in~\eqref{34b} to obtain
\begin{equation}\label{pc2}
r_c=r_++\frac{Mr_+-\eta_2Q^2}{r_+^2-\eta_2Q^2}~\frac{r_+}{2}~(1-\omega) +\cdots,
\end{equation}
where we have used~\eqref{rp} and~\eqref{30}. For $\bar{r}_c$ we obtain
\begin{equation}\label{pc3}
\bar{r}_c=\frac{\eta_2Q^2}{r_+}+\frac{\eta_2Q^2(r_+-M)}{2(r_+^2-\eta_2Q^2)}~(1-\omega) +\cdots.
\end{equation}
In the physical case $M^2\geq Q^2$ and $\eta_2=+1$, $r_c$ approaches $r_+$ from above in the limit $\omega\to 1^{-}$, since in this case $r_+>M\geq |Q|$. This remains true in the case $\eta_2=-1$ for all $M^2$ and $Q^2$.

At the other limit, $\omega\to 0^{+}$, $r_c\to +\infty$.

Now, differentiating both sides of~\eqref{34b} with respect to $\omega$ we obtain
\begin{equation}\label{pc4}
\partial_{\omega}r_c=\frac{r_+(Mr_c-\eta_2Q^2)}{-\omega\sqrt{[(1+3 \omega ) M r_+ +2\eta_1\eta_2 Q^2 \lambda ^2 \omega]^2- 8\eta_2 \omega  (1+\omega ) Q^2 r_+^2}}.
\end{equation}
For the parameter space ($M,\,Q,\,\gamma,\,\eta_2,\,\omega$), where the square root is real, $\partial_{\omega}r_c$ is certainly negative and so $r_c$ decreases from $+\infty$ to $r_+$ as $\omega$ runs from $0^+$ to 1. This is obvious for $\eta_2=-1$. For $\eta_2=+1$, had we assumed that $r_c$ first decreased, reached a minimum value,  then increased again, we would obtain the minimum at $r_c=\eta_2Q^2/M<M<r_+$ for some value of $0<\omega<1$. But this is not possibe since $r_c$ must approach $r_+$ from above as $\omega\to 1^-$. This shows that for the parameter space ($M,\,Q,\,\gamma,\,\eta_2,\,\omega$), where the square root in~\eqref{pc4} and~\eqref{34c} is real, $r_c>r_+$ if $0<\omega<1$. In the same manner we can show that $\bar{r}_c<r_+$.

For $M^2\geq Q^2$, we have shown that a CP always exists for $0<\omega<1$. Thus, the flow of isothermal fluids onto normal or phantom EMD black holes (ultra-relativistic fluids $\omega=1/2$, radiation fluids $\omega=1/3$, and sub-relativistic fluids $\omega=1/4$) is always critical. 

As is well known the accretion of ultra stiff fluids ($\omega=1$) onto ordinary matter is non-critical. We have extended this conclusion to accretion of ultra stiff fluids onto normal and phantom EMD black holes, since in this case the CP, $r_c=r_+$, is at the horizon position, so no critical behavior is observed in the outer region of the event horizon.

\subsection{The case \pmb{$M^2<Q^2$}}
\begin{figure*}[ht]
	\centering
	\includegraphics[width=0.47\textwidth]{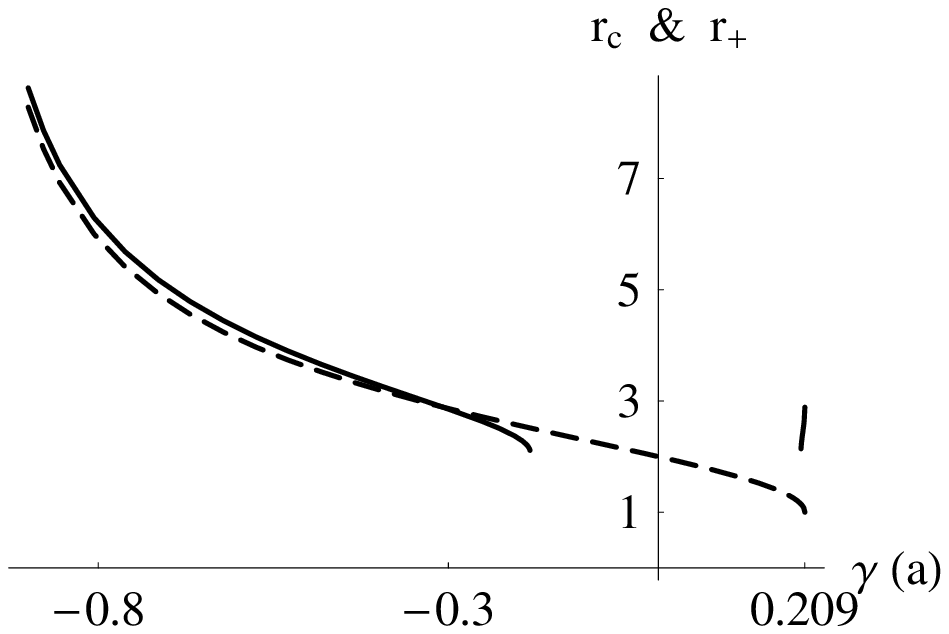} \includegraphics[width=0.47\textwidth]{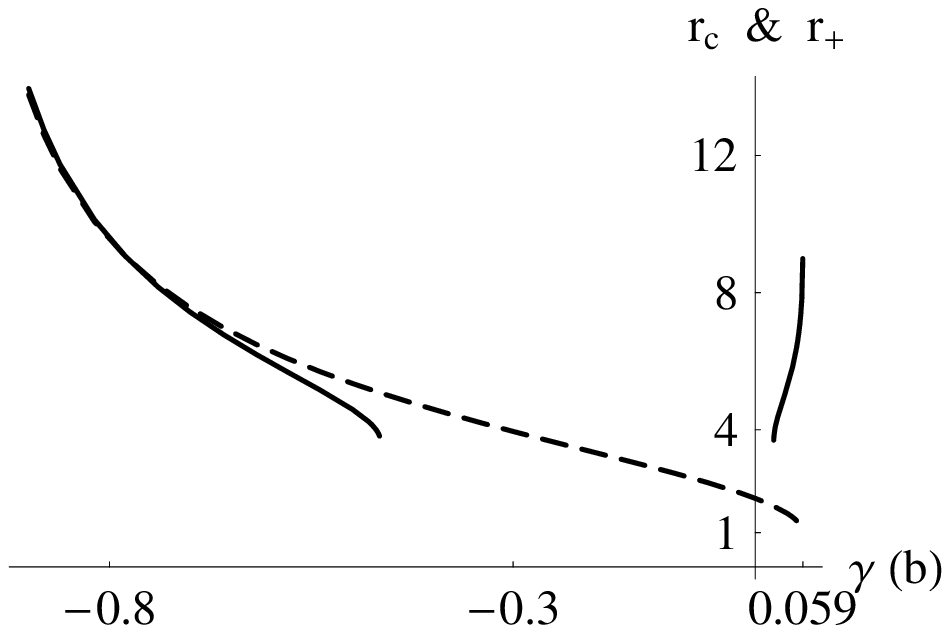}  \\
	\caption{\footnotesize{Plots of $r_c$~\eqref{34c} (continuous curve) and $r_+$~\eqref{30b} (dashed curve) vs. $\gamma$ for $\eta_1=\eta_2=+1$, $\omega=0.5$, $M=1$, and $M^2<Q^2$. The graph of $r_c$ is made of two disjoint curve segments, the one is monotonically decreasing and the other is monotonically increasing function of $\gamma$. (a): $|Q|=1.7$. (b): $|Q|=3$. It is clear from this figure that $r_+$, which is a monotonically decreasing function of $\gamma$, is not defined on the whole range of $\gamma$}~\eqref{gamma}; rather, it is defined for $-1<\gamma\leq \dfrac{M^2}{2Q^2-M^2}<1$. There are subintervals of $\gamma$ on which either $r_c<r_+$ or $r_c$ is not real. In these cases, the fluid accretion onto the corresponding black holes is non-critical.}\label{Fig1}
\end{figure*}

The EMD theory and its two derivatives E$\bar{\text{M}}$D and E$\bar{\text{M}}\bar{\text{D}}$ do admit black hole solutions with $M^2<Q^2$ (EM$\bar{\text{D}}$ has no black-hole solutions with $M^2<Q^2$). For fixed values of ($M,\,Q,\eta_1,\,\eta_2$) for which $r_+$ is real, the CP $r_c$ may not exist for some values of ($\gamma,\,\omega$) or may turn smaller than $r_+$ as shown in Fig.~\ref{Fig1}. In these cases, the fluid accretion onto the corresponding black holes is non-critical in the sense that~\eqref{d3} is satisfied but~\eqref{d4} is not; that is, as the fluid three velocity $v$ reaches the value $a$ of the three-dimensional speed of sound, during the accretion process, it does not do it in a stationary way so that $\dot{v}$ is nonzero. This means that, during the accretion process, the speed $v$ increases monotonically from $\sim 1/r^2$~\eqref{35c}, at spatial infinity, to 1, at the event horizon $r_+$. While in a critical flow, as $v$ reaches $a$ it remains stationary there for a while, then it increases again.

Figure.~\ref{Fig1} depicts the functions $r_c(\gamma)$ (continuous curve) and $r_+(\gamma)$ (dashed curve) for the other parameters held constant with $M^2<Q^2$. The graph of $r_c(\gamma)$ is made of two disjoint curve segments, the one is monotonically decreasing and the other is monotonically increasing function of $\gamma$. The one-segment graph of $r_+(\gamma)$ decreases monotonically.  $r_+(\gamma)$ is not defined on the whole range of $\gamma$~\eqref{gamma}; rather, it is defined for $-1<\gamma\leq \dfrac{M^2}{2Q^2-M^2}<1$.

We have obtained similar figures to Fig.~\ref{Fig1} for all $0<\omega<1$ and $M^2<Q^2$.

\section{Mass accretion rate and backreaction\label{secmar}}
The accretion rate is the change in the black-hole's mass per unit time. This is related to the flux of $T^{\mu\nu}$ by
\begin{equation}\label{mar1}
\dot M=-4\pi C(r) T^{\ r}_{t}(r)\big|_{r=r_+}.
\end{equation}
With $T^{\ r}_{t}=(e+p)uu_t$ it is easy to show, using Eqs.~\eqref{5} to~\eqref{11}, that $T^{\ r}_{t}=C_1C_2/(C\sqrt{D})$, consequently we have
\begin{equation}\label{mar2}
\dot M=-\frac{4\pi C_1C_2}{\sqrt{D(r)}|_{r=r_+}}.
\end{equation}
For most known black holes the function $D(r)\equiv 1$, as is the case with normal and phantom EMD black holes~\eqref{28}, so that the mass accretion rate reduces to
\begin{equation}\label{mar3}
\dot M=-4\pi C_1C_2.
\end{equation}
The values of the constants of motion ($C_1,\,C_2$) depend on the values of the enthalpy, number density and three-speed at spatial infinity. Using the same Eqs.~\eqref{5} to~\eqref{11} we obtain
\begin{align}
\label{mar4}&C_1=\frac{\sqrt{A(r)}C(r)n(r)v(r)}{\sqrt{1-v(r)^2}}\bigg|_{r\to\infty},\\
\label{mar5}&C_1C_2=\frac{A(r)C(r)n(r)h(r)v(r)}{1-v(r)^2}\bigg|_{r\to\infty}.
\end{align}
For an accretion process $v(r)$ and $C_1$ are negative. If the flowing fluid is made of ordinary matter, $h>0$ and thus $\dot M>0$ so that the mass of the black hole increases. Conversely, if the flowing fluid is made of phantom matter, $h<0$ and $\dot M<0$ and the mass of the black hole decreases.

If we momentarily restrict ourselves to normal and phantom EMD black holes and use Eqs.~\eqref{gi1} to~\eqref{gi3}, we obtain
\begin{equation}
\label{marh8}\dot M=4\pi v_{2}n_{\infty}h_{\infty},
\end{equation}
where $v_{2}$ and $n_{\infty}$ are positive and $h_{\infty}>0$ for ordinary fluids and $h_{\infty}<0$ for phantom fluids. As mentioned above, $n_{\infty}$ depends on the nature of the fluid where in many astrophysical applications it is taken as an ideal gas~\cite{inviscid}. The energy flux~\eqref{marh8} does not depend on the parameters of the EMD black hole ($M,\,Q,\,\gamma,\,\eta_2$), however, the whole process of accretion depends on them. For instance, one may seek to impose constraints on these parameters requiring that the accreting matter (the perfect fluid) have the same three-speed (at spatial infinity) to the third order in powers of $1/r$ and same particle density and enthalpy to the first order. This is the case when investigating the accretion of a given fluid and considering a set of different black holes. In this case we will have the following constraints on the parameters of these black holes: $M$ and the term $(1-\gamma)r_-$, which depends on the still-free parameters ($Q,\,\gamma,\,\eta_2$), must have the same values for the black holes.

Now, back to the most general case of accretion~\eqref{mar3}. If
\begin{equation*}
t\ll \tau_0\equiv \frac{M_i}{4\pi |C_1C_2|},
\end{equation*}
where $M_i$ is the initial mass of the black hole, to the first order in $t/\tau_0$ the backreaction effects of the fluid tend to modify the mass of the black hole according to the linear law
\begin{equation}\label{br1}
M=M_i-4\pi C_1C_2t+\cdots =M_i\Big[1-\text{sgn}(C_1C_2)\frac{t}{\tau_0}+\cdots\Big],
\end{equation} 
where $\text{sgn}(C_1C_2)$ is the sign of $C_1C_2$. Here $\tau_0$ is a characteristic time of accretion.

Note that both constants $C_1$ and $C_2$ in~\eqref{mar3} and~\eqref{br1} are independent of the mass of the black hole. Now, back to the steps leading to Eq.~\eqref{5}. Since $M$ is assumed constant in those steps one may {\textquotedblleft accidently\textquotedblright} multiply and divide the rhs of~\eqref{5} by $M^{\alpha}$, where $\alpha$ is some parameter (taken equal to 2 in Ref.~\cite{Babichev2}), to obtain
\begin{equation}\label{5b}
C\sqrt{D}nu=M^{\alpha}\mathcal{C}_{1}\quad\text{with}\quad D\equiv \frac{A}{B},
\end{equation}
where now the new {\textquotedblleft constant\textquotedblright} $\mathcal{C}_{1}\equiv C_{1}/M^{\alpha}$ depends on the mass of the black hole. This converts~\eqref{mar3} to
\begin{equation}\label{mar3b}
\dot M=-4\pi \mathcal{C}_{1}C_2~M^{\alpha}.
\end{equation}
As far as $4\pi \mathcal{C}_{1}C_2$ is seen as a constant, one may integrate~\eqref{mar3b}. For instance if $\alpha\neq 1$, one obtains
\begin{equation}\label{br2}
M=M_i\Big(1-\frac{t}{\tau}\Big)^{1/(1-\alpha)},
\end{equation}
where
\begin{equation}\label{br3}
\tau\equiv \frac{1}{4\pi \mathcal{C}_{1}C_2(1-\alpha)M_i^{\alpha-1}},
\end{equation}
could be positive or negative (had we taken $\alpha=1$, we would have obtained a log function in~\eqref{br2} and a different expression for $\tau$). Taking $\alpha=2$ in~\eqref{br2}, this reduces to Eq.~(7) of Ref.~\cite{Babichev2}. As far as backreaction effects are neglected, Eq.~\eqref{5b} is correct; however, when backreaction effects are taken into consideration Eq.~\eqref{mar3b} is correct to first order only since $\mathcal{C}_{1}$ depends on $M$. Thus, Eq.~\eqref{br2} is also valid to first order only and its series expansion in powers of $t$ reduces to the rhs of~\eqref{br1} once we replace $\mathcal{C}_{1}$ by $C_{1}/M^{\alpha}$. Its term in $t^2$, however, depends well on $\alpha$ even after replacing $\mathcal{C}_{1}$ by $C_{1}/M^{\alpha}$ and thus it does not produce the correct expansion term in $t^2/\tau_0^2$.

It is obvious from the analysis made here that the backreaction effects have been evaluated assuming a perturbed metric but a non-perturbed fluid~\cite{Babichev3}. In order to determine the term in $t^2/\tau_0^2$ in~\eqref{br1} one needs to consider a more advanced analysis where both the metric and the fluid are perturbed.

\section{Conclusion\label{seccon}}
The analysis made in this work is general and concerns accretion onto static solutions. It includes some of the spherical accretion work done previously on the Schwarzschild and Reissner-Nordstr\"om  black holes. Our model shares all known features with previously investigated accretion works. The distinguished features discovered in this work, which are emphasized in the two plots of Fig.~\ref{Fig1}, are characteristic of accretion on EMD black holes.

As far as the physical condition $M^2\geq Q^2$ is observed, the accretion process of uncharged ordinary isothermal fluids onto ordinary or phantom EMD black holes is characterized by the presence of a critical point $r_c$ through which the process is critical; in that, the three speed of the fluid becomes stationary for a while as it reaches the speed of sound. All accretion processes terminate at the horizon with the limiting speed of 1.

When the condition $M^2\geq Q^2$ is not observed, the accretion process onto some of the EMD black holes becomes non-critical.

In our investigation, we have restricted ourselves to isothermal fluids and we could extend the study, at least numerically, to other fluids. In the literature many authors have considered polytropic fluids but for such systems global solutions do not exist~\cite{global}. One can also assume a cosmological constant, vacuum energy or dark energy but their accretion does not have a substantial physical impact on the black hole.

Although in the literature, there are a lot of studies on this concept but still one can extend the analysis to include spinning black holes with non-adiabatic systems to take into account the terms of heat transport or viscosity.

\section*{Appendix: Reality of the roots of Eq.~\eqref{34b} for \pmb{$M^2\geq Q^2$} \label{secaa}}
\renewcommand{\theequation}{A.\arabic{equation}}
\setcounter{equation}{0}
\subsection*{Case \pmb{$\eta_2=-1$}}
For $\eta_2=-1$ the discriminant of~\eqref{34b}, which is the expression under the square root in~\eqref{34c}, is manifestly positive and so the roots $r_c$~\eqref{34c} and $\bar{r}_c$ are real.

\subsection*{Case \pmb{$\eta_2=+1$}}
The discriminant of~\eqref{34b} is positive or zero if
\begin{equation}\label{ap1}
(1+3 \omega ) M r_+ +2\eta_1\eta_2 Q^2 \lambda ^2 \omega\geq 2\sqrt{2}\sqrt{\omega(1+\omega)}|Q|r_+\qquad (\eta_2=+1,\ \eta_1=\pm 1).
\end{equation}

\subsubsection*{Subcase $\eta_2=+1$ and $\eta_1=+1$}
For $0<\omega\leq 1$ it is easy to show that $1+3 \omega\geq 2\sqrt{2}\sqrt{\omega(1+\omega)}$. Then, if $M^2\geq Q^2$ we will have $(1+3 \omega ) M r_+\geq 2\sqrt{2}\sqrt{\omega(1+\omega)}|Q|r_+$ and so~\eqref{ap1} is
trivially satisfied for $\eta_1=+1$.

\subsubsection*{Subcase $\eta_2=+1$ and $\eta_1=-1$}
Using~\eqref{rp} to express $Mr_+$ in terms of $r_+^2$ and $Q^2$ and the third expression in~\eqref{30} to express $\lambda^2$ in terms of $\gamma$~\eqref{gamma}, we bring~\eqref{ap1} to the form
\begin{equation}\label{ap2}
(\sqrt{2\omega}~r_+-\sqrt{\omega+1}~|Q|)^2+\dfrac{1-\omega}{2}\Big(r_+^2-\frac{2Q^2}{1+\gamma}\Big)\geq 0.
\end{equation}
Knowing that $0<\omega\leq 1$, this is trivially satisfied (for all $M^2$ and $Q^2$) if $\gamma<-1$~\eqref{gamma}. Now, if $\gamma\geq 1$~\eqref{gamma} and $M^2\geq Q^2$, since $r_+>M$~\eqref{30b}, we will have
\begin{equation}\label{ap3}
r_+^2-\frac{2Q^2}{1+\gamma}>M^2-\frac{2Q^2}{1+\gamma}\geq M^2-Q^2\geq 0,
\end{equation}
and so~\eqref{ap2} and\eqref{ap1} are satisfied.

\section*{Acknowledgments}
We thank Manuel E. Rodrigues for showing interest in an early stage of this work.


\begin{thebibliography}{99}
\bibitem{1} D.N. Spergel et al., WMAP Collaboration. Astrophys. J. Suppl. \textbf{170}, 377 (2007)
\bibitem{2} S. Perlmutter et al., Supernova Cosmology Project Collaboration. Astrophys. J. \textbf{517}, 565 (1999)
\bibitem{3} D.J. Eisenstein et al., SDSS Collaboration. Astrophys. J. \textbf{633}, 560 (2005)

\bibitem{Babichev} E.O. Babichev, V.I. Dokuchaev and Yu.N. Eroshenko, Phys. Usp. \textbf{56}, 1155 (2013)
\bibitem{4} R.R. Caldwell, Phys. Lett. B \textbf{545}, 23 (2002)
\bibitem{5} R.R. Caldwell, M. Kamionkowski, N.N. Weinberg, Phys, Rev. Lett. \textbf{91}, 071301 (2003)
\bibitem{6} S. Nojiri and S.D. Odintsov, Phys. Lett. B \textbf{562}, 147 (2003)
\bibitem{7} E.J. Copeland, M. Sami and S. Tsujikawa, Int. J. Mod. Phys. D \textbf{15}, 1753 (2006)
\bibitem{8} V. Sahni, A.A. Starobinsky, Int. J. Mod. Phys. D \textbf{9}, 373 (2000)

\bibitem{9} I.G. Martnez-Pas, T. Shahbaz, J.C. Velzquez, \textit{Accretion Processes in Astrophysics} (Cambridge University Press, 2014)
\bibitem{10} J. Karkowski, B. Kinasiewics, P. Mach, E. Malec and Z. Swierczynski, Phys. Rev. D. \textbf{73}, 021503 (2006)
\bibitem{11} P. Mach, E. Malec and J. Karkowski, Phys. Rev. D \textbf{88}, 084056 (2013)
\bibitem{12} P. Mach and E. Malec, Phys. Rev. D \textbf{91}, 124053 (2015)

\bibitem{Babichev2} E. Babichev, V. Dokuchaev and Yu. Eroshenko, Phys. Rev. Lett. \textbf{93}, 021102 (2004)

\bibitem{Jamil} M. Jamil, M. Rashid and A. Qadir, Eur. Phys. J. C \textbf{58}, 325 (2008)

\bibitem{super1} Z. Stuchl\'{i}k, Bull. Astron. Czechosl.  \textbf{31}, 129 (1980)
\bibitem{super2} M. Blaschke and Z. Stuchl\'{i}k, Phys. Rev. D \textbf{94}, 086006 (2016)

\bibitem{Bondi} H. Bondi, Mon. Not. R. Astron. Soc., \textbf{112}, 195 (1952)
\bibitem{Michel} F.C. Michel, Astrophys. Space Sci. \textbf{15}, 153 (1972)

\bibitem{Debnath} U. Debnath, Eur. Phys. J. C. \textbf{75}, 129 (2015)

\bibitem{A1} A.K. Ahmed, M. Azreg-A\"{\i}nou, M. Faizal and M. Jamil, Eur. Phys. J.C \textbf{76}, 280 (2016)
\bibitem{A2} A.K. Ahmed, M. Azreg-A\"{\i}nou, S. Bahamonde, S. Capozziello and M. Jamil Eur. Phys. J. C \textbf{76}, 269 (2016)
\bibitem{A3} M.~Azreg-A\"{\i}nou, Eur. Phys. J.C \textbf{77}, 36 (2017)
\bibitem{A4} A.K. Ahmed, U. Camci and M. Jamil, Class. and Quan. Grav. \textbf{33}, 215012 (2016)
\bibitem{A5} M.~Azreg-A\"{\i}nou, Phys. Rev. D \textbf{95}, 083002 (2017)


\bibitem{13} M. Sharif and G. Abbas, Mod. Phys. Lett. A \textbf{26}, 1731 (2011)
\bibitem{14} M. Sharif and G. Abbas, Chin. Phys. Lett. \textbf{28}, 090402 (2011)

\bibitem{Amani} A.R. Amani and H. Farahani, Int. J. Theor. Phys. \textbf{51}, 2943 (2012)

\bibitem{Rezolla} L. Rezzolla, O. Zanotti, \textit{Relativistic Hydrodynamics} (Oxford University Press, 2013)
\bibitem{b1} P. Crawford, I. Tereno, Gen. Relativ. Grav. \textbf{34}, 2075 (2002)
\bibitem{b2} G.F.R. Ellis, R. Maartens, M.A.H. MacCallum, \textit{Relativistic Cosmology} (Cambridge University Press, 2012)

\bibitem{d2} G. Cl\'ement, J.C. Fabris and M.E. Rodrigues, Phys. Rev. D \textbf{79}, 064021 (2009)
\bibitem{c1} M.E. Rodrigues and Z.A.A. Oporto, Phys. Rev. D \textbf{85}, 104022 (2012)
\bibitem{c2} A. Nakonieczna, M. Rogatko and R. Moderski, Phys. Rev. D \textbf{86}, 044043 (2012)

\bibitem{d3} M. Azreg-A\"{\i}nou, Phys. Rev D \textbf{87}, 024012 (2013)

\bibitem{inviscid}M. Kafatos and R. Yang,
Mon. Not. R. Astron. Soc. \textbf{268}, 925 (1994)

\bibitem{Babichev3} E. Babichev, V. Dokuchaev and Yu. Eroshenko, Class. Quantum Grav. \textbf{29}, 115002 (2012)

\bibitem{global} P. Mach, Phys. Rev. D \textbf{91}, 084016 (2015)

\end{thebibliography}
\end{document}